\documentclass[aps,pre,twocolumn]{revtex4-1}
\usepackage{amsmath}
\usepackage{amssymb}
\usepackage{bm}
\usepackage{graphicx}
\usepackage{hyperref}
\usepackage[usenames, dvipsnames]{color}
\usepackage[table]{xcolor}
\usepackage{comment}
\begin{document}
\author{Shikha Bhadoria}\email{shikha.bhadoria@mpi-hd.mpg.de}
\author{Naveen Kumar} \email{kumar@mpi-hd.mpg.de}
\affiliation{Max-Planck-Institut f\"ur Kernphysik, Saupfercheckweg 1, 69117 Heidelberg, Germany}

\title{Collisionless shock acceleration of quasi-monoenergetic ions in ultra-relativistic regime}

\begin{abstract}
Collisionless shock acceleration of carbon ions (C$^{6+}$) is investigated in the ultra-relativistic regime of laser-plasma interaction by accounting for the radiation reaction force and the pair production in particle-in-cell simulations. Both radiation reaction force and pair plasma formation tend to slow down the shock velocity, reducing the energy of the accelerated ions, albeit extending the time scales of the acceleration process. Slab plasma target achieves lower energy spread while target with a tailored density profile, yields higher ion acceleration energies.
\end{abstract}

\maketitle

\section{INTRODUCTION}
Laser-driven ion acceleration from thin-foil targets is a promising area of research, having applications from proton radiography, fusion research~\cite{Macchi2013a,Daido:2012aa,Roth:2001aa,Fernandez:2009aa}, and especially cancer therapy~\cite{Ute:2011aa,*Linz:2016aa}. Laser-driven ion beams have short duration ($\sim$fs), high currents ($\sim$MA), and low emittance. It makes them a unique tool to study fundamental science such as warm-dense matter in a laboratory~\cite{Patel:2003aa}.

Laser-driven-ion acceleration employs several different schemes for accelerating the ions \emph{e.g.} target normal sheath acceleration (TNSA)~\cite{Wilks:2001aa,*Schreiber:2006aa}, radiation pressure acceleration (RPA)~\cite{Esirkepov:2004aa,Qiao:2012aa,Chen:2009aa,Dollar:2012aa,Tripathi:2009aa,Schlegel:2009aa,Macchi:2010aa,Chen:2009aa,Henig:2009aa,Weng:2012aa,Tamburini:2010aa,Capdessus2015a,Wu:2015aa}, direct laser acceleration~\cite{Salamin:2008aa}, breakout-after-burner (BOA)~\cite{YIN:2006aa,*Albright:2007aa}, and collisionless shock acceleration (CSA)~\cite{Silva:2004aa,Fiuza:2012ab,Haberberger:2011aa,Chen:2007aa,Shikha-Bhadoria:aa} etc. TNSA was the first mechanism proposed for laser-driven ion acceleration and it works in every interaction scenario. The spectrum of the ions accelerated by the TNSA is a Maxwellian which isn't desirable for cancer therapy. The RPA mechanism was suggested to be extremely efficient both for maximum ion energy and the spectrum quality in the ultra-relativistic regime ($I_l\gg 10^{21}$W/cm$^2$). The efficacy of the RPA scheme is yet to be proven experimentally and it is expected to work best for the high-contrast laser pulse and ultra-thin ($\sim $nm thickness) targets. The BOA scheme operates in the relativistic transparency regime where the laser pulse can penetrate the target. The distinct feature of the BOA scheme is the onset of the relativistic Buneman instability which facilitates the transfer of energy from electrons to ions~\cite{YIN:2006aa,*Albright:2007aa}. Ion acceleration from relativistic transparency regime has been experimentally demonstrated in the interaction of an intense laser with ultra-thin targets~\cite{Palaniyappan:2015aa}. Incidentally, in this scenario, the RPA mechanism can also be dominant. Because of this reason, there appears to be some overlap in defining the dominant mechanisms of the ion acceleration from an ultra-relativistic laser interacting with ultra thin-foil targets. 

The CSA scheme (works best for $\sim \mu$m thick targets) has emerged as an alternative scheme which has the potential to produce a high-quality ion beam~\cite{Haberberger:2011aa,Pak:2018aa}, if the competing TNSA field is controlled by tailoring of the target~\cite{Fiuza:2012ab}. This scheme has also attracted significant attention as the formation dynamics of the collisionless shocks in laser-plasma interaction constitutes an important part of the newly developing area of research known as the laboratory astrophysics~\cite{Fiuza:2012aa}. From the point of view of the laser-driven ion acceleration, the ion acceleration from electrostatic shocks in near-critical density plasmas is shown to provide a high-energy ion beam with a low energy spread~\cite{Silva:2004aa,Fiuza:2013aa,Haberberger:2011aa,Pak:2018aa}. 

Though the potential of aforementioned schemes for the ion-acceleration seems to be promising, the stated goal of the tumour therapy $\sim$ (120-430) MeV/u for high-Z target \cite{Ute:2011aa,*Linz:2016aa,Macchi2013a} is yet to be demonstrated experimentally. Currently this is, in large part, due to the unavailability of many ultra-intense laser systems that can deliver intensity $I_l\gg10^{22}$W/cm$^2$. But this is bound to change soon due to various  laser-systems upgrades planned around the world~\cite{ELI:aa,Vulcan:aa,XCELS:aa}. In this ultra-relativistic regime where effects like radiation reaction and pair production become important, theoretical studies have largely focused on the RPA mechanism due to reasons discussed before. Ion acceleration in the relativistic transparency regime has also been studied~\cite{Henig:2009aa,Tamburini:2010aa,Palaniyappan:2012aa,Palaniyappan:2015aa,Yin:2011aa,Kim:2015aa,Weng:2012aa,Capdessus2015a,Wu:2015aa}. However, the CSA of ions has not been investigated in this regime until now.

In this paper, we study the laser-driven CSA of ions from the near-critical-density (NCD) targets in the ultra-relativistic regime by including the effects of radiation reaction (RR) force and pair production (PP) due to the Breit-Wheeler (BW) process. We consider both slab as well as tailored NCD targets in particle-in-cell (PIC) simulations. In Sec.\ref{slab_target}, we show  shock structure formations in all three cases viz. no RR force, RR force, and RR + PP using PIC simulations. The inclusion of the RR force and PP in the plasma dynamics lower the piston velocity and consequently the shock velocity, leading to the lower ion energy gain in both cases. To further optimize the CSA of ions, we also present, in Sec.\ref{tailored_target}, results from a tailored plasma target~\cite{Fiuza:2012ab}.

\section{CSA of ions for a thick plasma slab target}\label{slab_target}

We carry out 2D PIC simulations using the open-source code EPOCH~\cite{Arber:2015aa}. EPOCH includes quantum RR force and PP by the probabilistic Monte-Carlo method. We employ a linearly polarised laser pulse, impinging on the left boundary with a finite temporal profile $I(t)= I_0 \exp [-(t-t')^2/\tau_0^2]$, with $\tau_0=400$fs and $t'=200$ fs. The laser peak intensity is $I_0 = 1.2386 \times 10^{23}$ W/cm$^2$, ($a_0=e E_L/m_e\omega c =300 $), where $m_e$ is the electronic mass, $e$ is the electronic charge, $\omega$ is the laser carrier frequency, $E_L$ is the electric field of the laser and $c$ is the velocity of the light in vacuum. It interacts with a pre-formed fully-ionised carbon plasma (C$^{6+}$) with a temperature $T_{e^-}=T_{\text{C}^{6+}}=700$ eV and electronic density, $n_{e}=300n_c$, where $n_c = m_e \omega^2 / 4 \pi e^2$ is the non-relativistic critical density of a plasma for $1\,\mu$m laser wavelength. The target has a thickness of $40\, \mu$m, and is located at $6\, \mu$m from left boundary of the simulation box. We employ transmitting and periodic boundary conditions in $x$ and $y$ directions, respectively. The simulation box has dimensions of $L_x \times L_y = (150\mu \textrm{m} \times 6\mu \textrm{m}$), with the cell size: $\Delta_x \times \Delta_y = (10 \textrm{nm} \times 10 \textrm{nm}$) and uses $50$ particles per cell.

\subsection{Shock structures in three cases}

\begin{figure}
\centering
\includegraphics[height=0.38\textheight,width=0.52\textwidth]{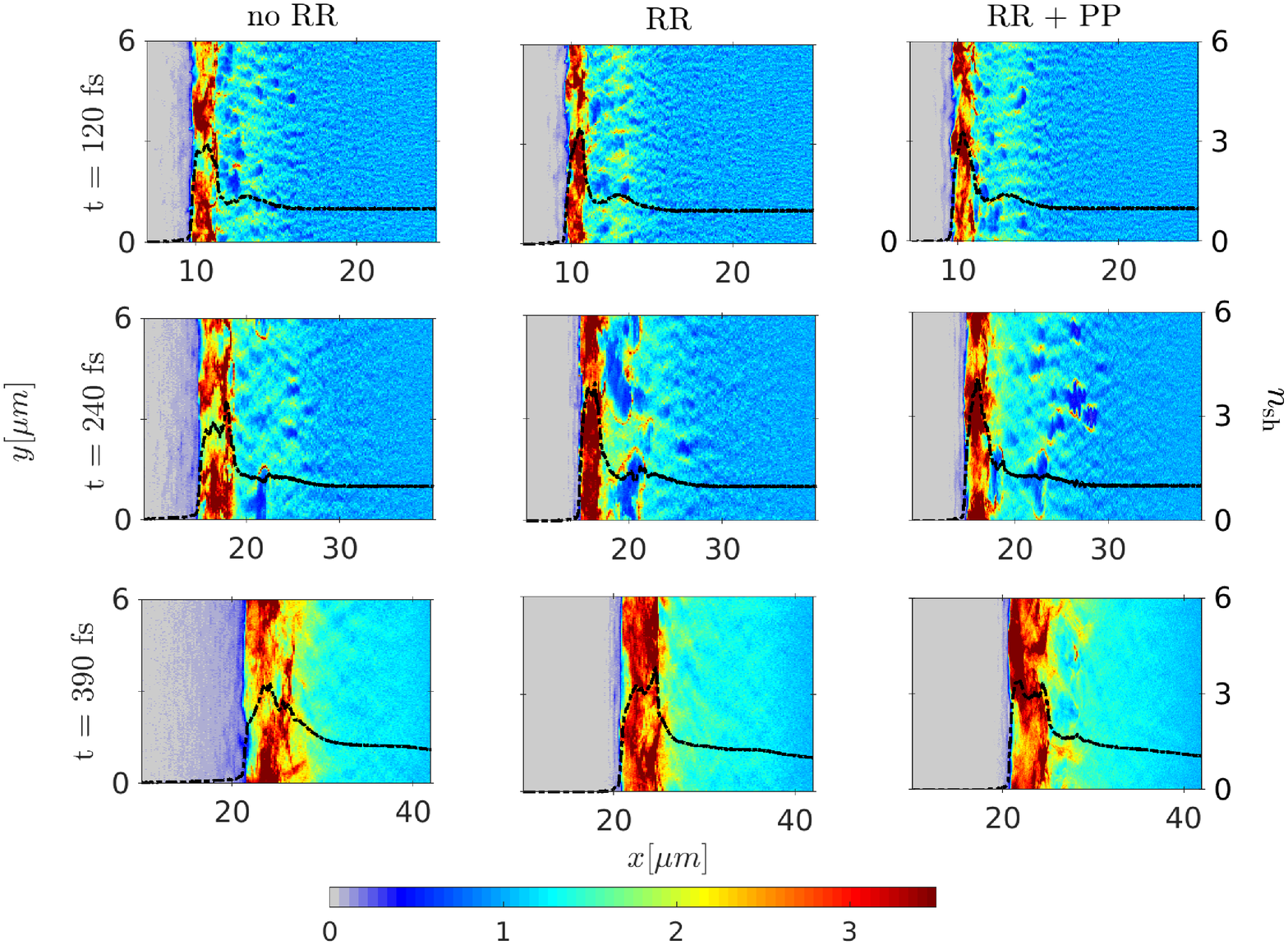}
\caption{Number density of plasma normalised by initial density at different instants. First, second and third columns represent the cases of no RR, with RR and with RR + PP respectively. The shock density jump, $n_{\textrm{sh}}$ (second vertical axis), averaged in the $y$-direction is over-plotted with a black dotted line in each case.}
\label{fig1}
\end{figure}

\begin{figure}
\includegraphics[height=0.34\textheight,width=0.55\textwidth]{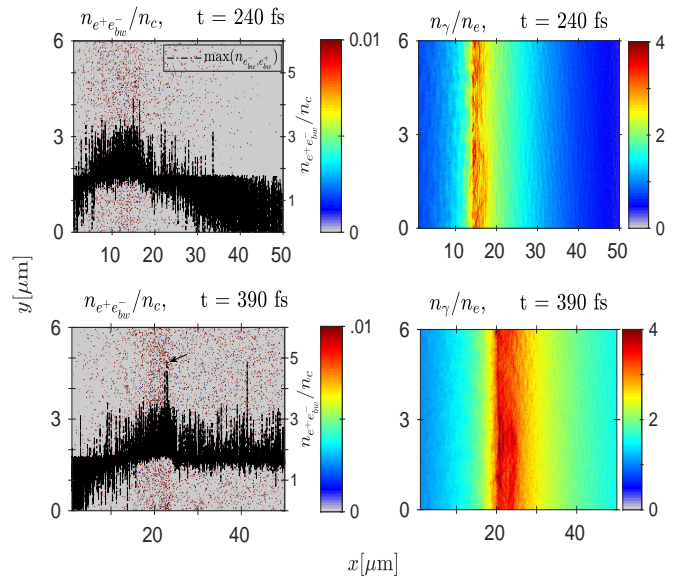}
\caption{Densities of pairs (left column, normalised by $n_c$) and the photons (right column, normalized by initial electron density, $300 n_c$) generated at different times corresponding to the last column in Fig.\ref{fig1}. The peak pair density (black) is over-plotted in the first column. The peak value reaches upto $\sim 5n_c$ (incidentally matching with the y-axis) around the laser plasma interface at 390 fs (marked by an arrow in the corresponding panel in second row).}
\label{fig2}
\end{figure}
 
Fig.\ref{fig1} shows the temporal evolution of the shock structure in three cases (no RR, RR and RR+PP) at different instants. One can clearly see a shock structure formation with a density jump $n_{\textrm{sh}}= n_d/n_u \sim 3$ , where $n_d$ and $n_u$ are the densities of the downstream and upstream plasmas respectively, in all three cases at $t=120$ fs (first row). To the right of the target one may also see filamentary structures. These structures are generated when the hot-electrons (generated at the laser-plasma interface) traverse the target and excite a cold return plasma current, leading to filamentation due to the Weibel instability. Later on (second row), one sees that the laser-plasma interface starts to become relativistically transparent leading to a deeper laser penetration and a stronger target heating in the first case. The target mass to the left of the laser-plasma interface 
 (mainly hot electrons) is suppressed in the case of RR (second column) and RR+PP (third column). Widening of the shock and suppression of the target mass to the left are due to the RR force induced plasma dynamics. Due to the RR force, the electrons experience significant energy losses limiting their excursion in the target. The effect of RR force is stronger for the electron moving towards the laser, hence, a significant suppression of the target mass to the left of the target is expected. One may also notice that the laser penetration inside the target is reduced in the last two columns, as expected, since a significant fraction of laser energy is lost as radiations. Combined effects of smaller electron excursion and energy loss result in a stable but wider shock front  as seen in Fig.\ref{fig1}. One may also note that the results in the second and third columns (Fig.\ref{fig1}) are almost identical, suggesting no stronger effect of the pair production on the shock structure.  At $t=390$ fs, (accounting for RR force and PP), the shock structure remains uniform and has a smooth density variation across its width. It is also interesting to see that the density compression of the shock, $n_{sh}$, is increased ($\sim 10-15\%$) in the second and third column where the QED effects (RR+PP) are taken into account.

\subsection{Particles generations and energy partitioning}

As noticed in Fig.\ref{fig1}, the effect of PP on the shock structure is minimum, one can expect smaller density of pairs generated in our case. Fig.\ref{fig2} shows the densities of the photons emitted (right column) and pairs produced by the BW process (left column) at two different instants corresponding to the last two rows of Fig.\ref{fig1}. The peak density of pairs (in black) is overplotted in the first column of Fig.~\ref{fig2}. As expected, densities of the photons and pairs increase with time. At an earlier instant ($t=240$ fs), density of the pairs produced is rather low while there is a copious amount of the high energy photons ($E_{\gamma}\ge 2$ MeV) generated ($ 3-4 n_e $) at $t=240$ fs (upper row). Both are concentrated at the laser-target interface. In EPOCH PIC simulations, one can distinguish the electrons generated by the BW process from the background plasma electrons. The pair density peaks to 
 $\sim 5 n_c$ at the laser plasma interface, which can be large enough for raising the threshold for further pair-production due to enhanced screening of the laser field at the target surface, as also envisaged in the pair plasma-cushion scenario~\cite{Kirk:2013aa}. Consequently, the laser penetration into the target is weaker in the third case (RR + PP). 

\noindent
\begin{figure}
\centering
\includegraphics[height=0.37\textheight,width=0.5\textwidth]{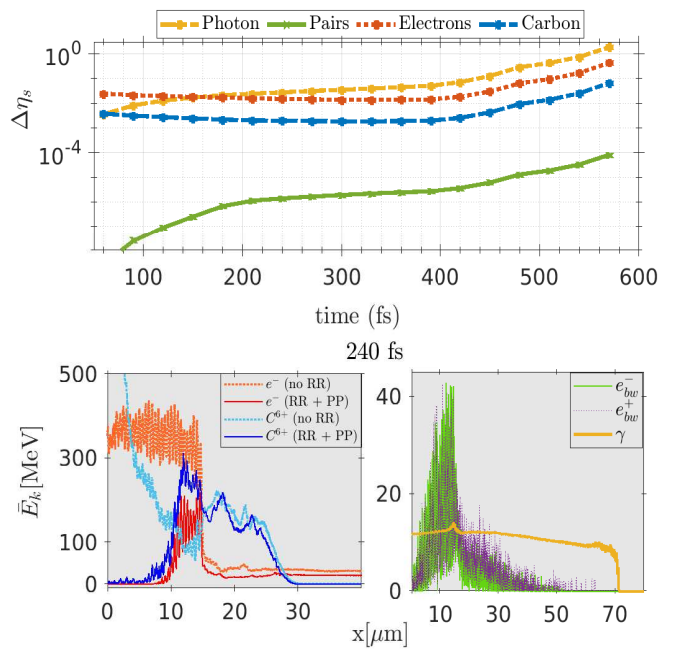}
\caption{Fraction of average laser energy $ \mathcal{E}_{L} $ being converted to each species \emph{`s'} ($\Delta \eta_s = \mathcal{E}_{s}/\mathcal{E}_{L}$) in logarithmic scale (top panel). The BW electrons (green lines in upper panel) are distinguished from the plasma electrons (red line in upper panel). Average kinetic energy ($y$-averaged) for each species at $t=240$fs (lower panel). The electron and ion energies are shown in the lower left panel, while the pairs and photons are shown in the lower right panel. }
\label{fig3}
\end{figure}
\noindent
Fig.\ref{fig3} (top panel) depicts the re-partitioning of the laser energy  among different species in the simulation ($\Delta \eta =\mathcal{E}_{s}/\mathcal{E}_{L}$, where $\mathcal{E}_{s}$ is the energy per unit length of each species s' and $\mathcal{E}_{L}$ is that of the laser pulse). Here, $ \mathcal{E}_{L} (t) = \int_0^{L_x}\int_0^{L_y} (1/8\pi)\{ E_y ^2 + B_z^2 \}\,dx\, dy\,$, is the instantaneous laser energy per unit length, where $L_x$ and $L_y$ are the dimensions of the simulation box. $\mathcal{E}_{s}$ is calculated by integrating the energy distribution of each species at a given instant from the PIC simulation. As one can expect, this ratio increases with time for each species and then saturates between $200$ to $400$ fs. At $400$ fs, the laser pulse acquires the maximum amplitude and the energy, consequently this ratio begins to increase. Since in this case, actual laser energy begins to decrease while the fraction gone into the photon energy remains constant. At $t=400$ fs, the fraction of laser energy gone into the photons is $\Delta\eta_{\gamma} \sim 15 \%$, consistent with Ref.~\cite{Brady:2012aa}. However, the fraction of laser energy is converted into producing pairs is significantly lower, as expected from Fig.\ref{fig2}. To exclude the influence of multi-dimensional effects on the energy conversion process, we also show in Fig.\ref{fig3} (lower panel), the spatial distribution of the $y$-averaged kinetic energy for each species (plasma electrons and ions are shown in the lower left panel, whereas the photons and pairs are depicted in the lower right column)  at $t=240$ fs.  For comparison, energies of the electrons and carbon ions in the case of no RR force are also overplotted with dotted lines in the lower left column.  One can see that there is a significant difference between the spatial distribution of electron's kinetic energy in the case of with and without RR force. The plasma ions and electrons form a double-layer structure.  RR force significantly suppresses the energy of the electrons  counter-propagating to the laser pulse as seen in the bottom left panel of Fig.\ref{fig3} due to radiative cooling. One can also see (lower right panel) that the kinetic energies of the pairs (electrons and positrons) produced by BW process are similar to  each other, as expected. The BW positrons  show slightly higher energy gain than the BW electrons. This is due to the space-charge field that pulls the BW-electrons while it repels the BW-positrons in the regions of charge imbalance in the plasma. Consequently, the BW positrons can gain more energy than the BW electrons in the laser field. A small bump in the average kinetic energy of the photons (lower right panel) corresponds to the laser plasma interface where radiation emission is maximum. Though BW electrons and positrons are generated by the interaction of the high-energy photons ($\sim 15$ MeV) with the laser pulse, the generated pairs can further be accelerated by the laser field and consequently attain energies higher than the photons as seen in Fig.\ref{fig3} (bottom right panel). 

The partitioning of the laser energy among different species leads to the slowing down of the laser-piston and consequently lower shock velocities. 
One can estimate the lowering of the piston velocity in each case by employing the energy and momentum flux conservations, which reads as,
\begin{equation}
  (1 - \mathcal{R}) (1-\beta_p) =\Delta \eta_{\gamma} + 2 \frac{\gamma_p ^2 \beta_p^3}{B^2}  + (\gamma_h -1) \frac{ m_e n_{he}}{B^2 M n_i},
  \label{eq1}
\end{equation}     
\begin{equation}
  (1 + \mathcal{R}) (1-\beta_p)  = \mathcal{P_{\gamma}} + 2 \frac{\gamma_p ^2 \beta_p^2}{B^2} + \gamma_h  \frac{ m_e n_{he}}{B^2 M n_i},
  \label{eq2}
\end{equation}      
where $\mathcal{R, P_{\gamma}},\Delta \eta_{\gamma}, n_{he},\gamma_h$ are reflection coefficient of the laser pulse, pressure of the emitted radiation, fraction of laser energy lost as radiations, hot electron's density and it's Lorentz factor, respectively, $M=Zm_e +m_i$ and $ B= a_0 \sqrt{n_c m_e / (2 n_i M) }$. One may note here that previous estimates of the slow down of the piston velocity were either attributed to the hot electron generation or to the high energy photons production respectively~\cite{Levy2013,Capdessus2015a,Nerush:2015aa}, while we consider the effect of both in our case. Solving for $\mathcal{R}$ gives, $\mathcal{R} =  \{(1-\beta_p)/(1+\beta_p) \} + \{(\gamma_h n_{he} m_e)/(2B^2Mn_i(1-\beta_p))\} -\{ (\Delta\eta_\gamma-\mathcal{P_{\gamma}}) /(2(1-\beta_p))  \}$. 
\noindent
\begin{figure}
\includegraphics[height=0.2\textheight,width=0.45\textwidth]{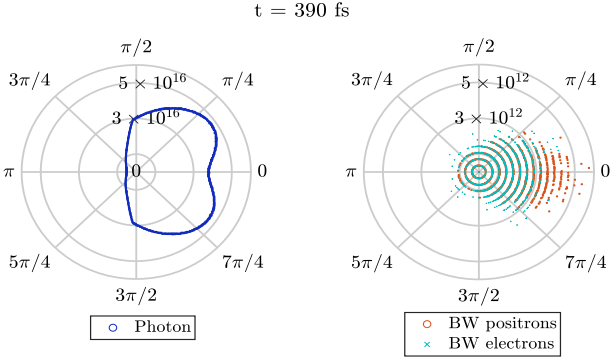}
\caption{Angular distribution of photons (left) and pairs (right) at 390 fs.}
\label{figangle}
\end{figure}
Substituting $\mathcal{R}$ in Eq.~\eqref{eq2}, yields 
\begin{equation}\label{eq3}
\beta_p^2  \alpha_0 + \alpha_1 \beta_p +\alpha_2  =0,
\end{equation}
where $\alpha_0= 4- (4+\Delta \eta_{\gamma} + \mathcal{P_{\gamma}} )B^2 - (2 \gamma_h-1) { m_e n_{he}}/{( M n_i)}$,  $\alpha_1=8B^2$, and $\alpha_2=(\Delta\eta_{\gamma}+ \mathcal{P_{\gamma}}-4 )B^2 + (2 \gamma_h-1) { m_e n_{he}}/{( M n_i)}$. The hot-electron density can be approximated as $n_{he} = 1.7 \times 10^{22} \rm{cm}^{-3} \times I_{18}^2 \times ({\tau_o}/{1\,\rm{ps}})\times({T_{he}}/{200\,\rm{keV}})^{-3} \times({\sigma_e}/{10^6\,\Omega^{-1}\,\rm{m^{-1}}})^{-1}$ where $I_{18} = I_{\rm{abs}} / 10^{18}$W/cm$^2$, $I_{abs}=0.8I_0$, $\tau_0$ is the laser period, $T_{he}$ is hot electron temperature, $\sigma_e$ is Spitzer conductivity of the target~\cite{Bell:1997aa}.  The radiation pressure due to photon emission, $\mathcal{P}_\gamma =\Delta \eta_{\gamma}\cos\langle \theta \rangle $ depends on the fraction of the laser energy converted into photons, $\Delta \eta_{\gamma}$, and the half-angle ($\theta$) of the circular cone in which the dominant photon emission occurs. Solving for $\beta_p$ numerically we find a good agreement between the calculations and the values from the PIC simulations as shown in Table~\ref{table_v}. Here, we take $\Delta \eta_{\gamma}=0.15$ and the average angle of radiation emission to be $ \langle \theta \rangle \sim 35^\circ$, as can be inferred from Figs.~\ref{fig3} and ~\ref{figangle} respectively. The BW positrons  density distribution in Fig.~\ref{figangle} (right panel) is identical to the BW electrons density distribution except there are slightly higher number of positron than the electrons along $\theta=0$. This is presumably due to the space-charge field that pulls the BW-electrons while repels the BW-positrons in the regions of charge imbalance in the plasma, consistent with the observation in Fig.\ref{fig3} (lower right panel).
\noindent
\begin{figure}
\centering
\includegraphics[height=0.35\textheight,width=0.48\textwidth]{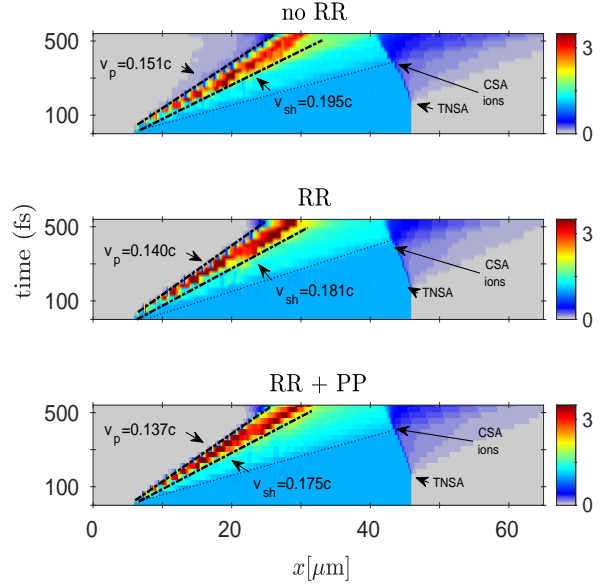}
\caption{$(x$-$t)$ plot of ion density indicating shock and  piston velocities in the three cases considered.}
\label{fig_xt}
\end{figure} 
\noindent
\begin{table}[b]
\begin{ruledtabular}
\begin{tabular}{cc|c|c|ccc|c}
&Cases & $\beta_p$ & $\beta_p$ & &$\beta_{sh}$\text{ (theo.)} & & $\beta_{sh}$ \\
&   & \text{(theo.)}&\text{ (sim.)}&$\Gamma_{ad}=2$&$\Gamma_{ad}=\frac{5}{3}$&$\Gamma_{ad}=\frac{4}{3}$&\text{ (sim.)} \\ \hline \\[-7pt]
&no RR  & 0.149 & 0.151 & 0.223  & 0.198 & 0.174  & 0.195 \\
&RR     & 0.141 & 0.140 & 0.211  & 0.188 & 0.164  & 0.181 \\
&RR+PP  & 0.140 & 0.137 & 0.210  & 0.187 & 0.163  & 0.175 \\
\end{tabular}
\end{ruledtabular}
\caption{\label{table_v}%
The normalised shock and piston velocities ($\beta_{p(sh)} \equiv v_{p(sh)}/c$) around $t=400$ fs for the three simulation runs shown in Fig.\ref{fig1}. The theoretical estimates are from Eq.\eqref{eq3} and PIC simulation results are inferred from the ($x$-$t$) plot in Fig.\ref{fig_xt}.
}
\end{table} 
\noindent
The piston and shock velocities (simulation) are computed from ($x$-$t$) plot shown in Fig.\ref{fig_xt}. For calculating the piston velocity in the RR+PP case, a higher value of the electron density was used ($\sim 305 n_c$), due to the pair production (see Fig.\ref{fig2}).  The theoretical estimate of shock velocity $v_{sh}$ for a non-relativistic hydrodynamic shock is related to the piston velocity $v_p$ as $v_{sh}=v_p (\Gamma_{ad}+1) /2$, where $\Gamma_{ad}$ is the adiabatic index of the plasma fluid~\cite{Ruyer:2015aa}. Hence, a reduction in the piston velocity implies  a similar reduction in the shock velocity as summarized in Table~\ref{table_v} for each case. For calculating the shock velocity, we employ different adiabatic indices. For a 2D simulation, the adiabatic index is usually taken as $\Gamma=2$ but the radiation generation and relativistic effects can lower it~\cite{Blandford:1976aa,Stockem:2012aa}. As one can see from the Table~\ref{table_v}, the adiabatic index $\Gamma=5/3$, for an ideal plasma fluid, shows a good agreement in the first two cases (no RR and RR). However, in the case of RR+PP, both adiabatic indices $\Gamma=5/3,\,\textrm{and}\, 4/3$, show larger deviations from the simulation value. The adiabatic index, $\Gamma=4/3$, valid for ultra-relativistic plasma always gives value slightly smaller than PIC simulation while adiabatic index $\Gamma=5/3$ gives slightly larger value. The slight deviations between the theoretical estimates and PIC simulation results of the piston velocities can be attributed to approximate estimation of the hot-electron energy and density in our calculations, especially in the later cases of RR and RR+PP. The slow-down of the shock velocity, on account of the energy partitioning in the last two cases is clearly visible.

\subsection{Electromagnetic field energy development, electron-ion phase spaces and ion energy spectra}\label{slab_spectrum}

\begin{figure}
\centering
\includegraphics[height=0.32\textheight,width=0.53\textwidth]{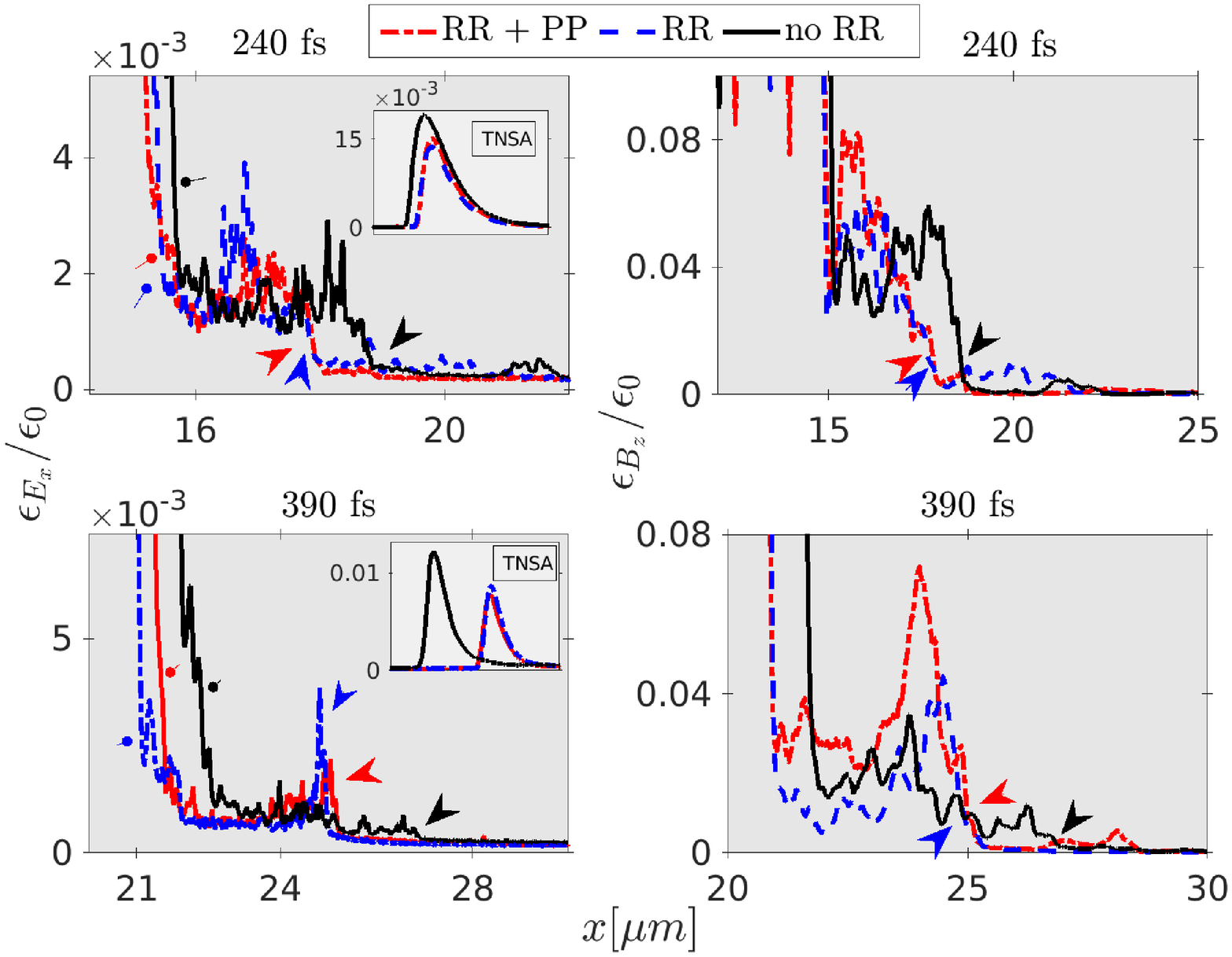}
\caption{Development of $y$- averaged electrostatic field energy densities due to $E_x$ (left column) and due to $B_z$ (right column) normalised by $(1/8\pi) E_L^2$  at t=240 fs and 390 fs. The shock front is marked by arrows of respective colors whereas, the position of laser piston is marked by the circular marker. The inset on the top in the left column shows the TNSA field at the end of the target.}
\label{fig_em}
\end{figure}

To discern the effects of the RR force and PP on the electromagnetic energies development due to the Weibel/filamentation instability, we plot in Fig.\ref{fig_em}, y-averaged energy densities associated with the longitudinal electric [$\langle E_x^2 \rangle_y / 8\pi $, first column] and the transverse magnetic fields [$\langle B_z^2\rangle_y / 8\pi $, second column]. One can attribute changes in the longitudinal electric field energy to energy re-partitioning  on electrostatic shock evolution dynamics. While the evolution of the transverse magnetic field energy is intimately connected to the Weibel/filamentation instabilities of the return plasma current, since these instabilities can generate an ultra-strong quasi-static magnetic field. One can see two interesting points here: first RR force and also PP increase the magnitudes of these field energies. Second, the shock front  moves faster without RR force and PP, as also seen in Fig.\ref{fig1}. Since RR force and PP re-partition the laser energy into different particles \emph{e.g.} electrons, ions, pair-plasma, and photons (as seen in the upper panel of Fig.\ref{fig3}), it slows down the hole-boring velocity and consequently the shock velocity in both cases; the reduction being ($\sim 10\%$) in the pair-plasma formation scenario, as can be expected. This can be seen in Fig.\ref{fig_xt} (and also Table~\ref{table_v}) where the shock velocities and piston velocities for all three cases have been calculated from the $(x$-$t)$ plot of ion density. The modification in the transverse magnetic field energy due to RR force is particularly interesting, suggesting that RR force (and PP) can also affect the magnetic field energy development due to the Weibel/filamentation instability.  In our case, RR force and pair-plasma formation affect the hot electron generation at the target surface. Since these hot electrons while traversing through the target excite the return current causing the onset of the Weibel/filamentation instability, the RR force and the pair-plasma affect the growth of the Weibel/filamentation instability in an indirect way. Though the effect of the RR force on the ion Weibel instability and the Weibel instability in counterpropagating pair-plasmas has been studied before~\cite{Grassi:2017aa,DAngelo:2015aa}, we have here shown the influence of RR force and  PP on the growth of the electron Weibel/filamentation instability in laser-plasma interaction.

\begin{figure}
\includegraphics[height=0.28\textheight,width=0.46\textwidth]{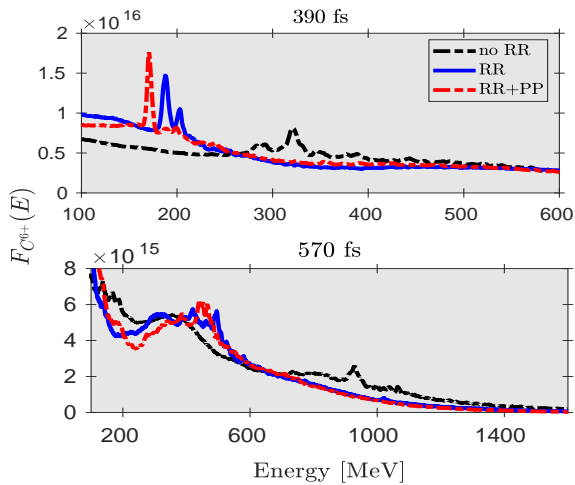}
\caption{Ion energy spectrum at $t=390$ and $570$ fs. The case without the radiation reaction force (black line) has a peak energy of 321 MeV with an energy spread of $\sim 3\%$ while for the other two cases (RR and RR + PP) it is 188 MeV and 170 MeV respectively with similar energy spreads $\sim 2-3\%$. At later time of 570 fs (bottom panel) it can be seen that the ions gain more energy (928 MeV with $\sim 2.6\%$ spread, 495MeV with $\sim 2.8\%$ spread, 465MeV with $\sim 1.7\%$ spread respectively).}
\label{fig_fe}
\end{figure}

\begin{figure}
\includegraphics[height=0.3\textheight,width=0.5\textwidth]{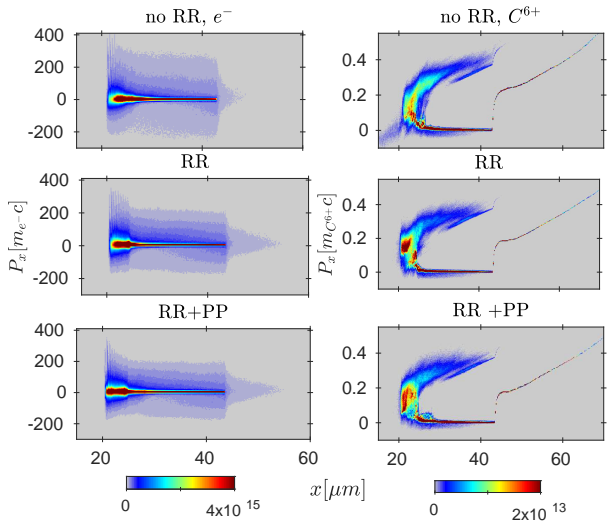}
\caption{Electron phase space (left column) and ion phase space (right column) at $t = 390 $ fs for all 3 cases}
\label{fig_xpx}
\end{figure}

The reduction of shock velocity indeed affects the maximum energy gain of ions as shown in Fig.\ref{fig_fe} (top panel), where the ion energy spectrum in each case is plotted at $t=390$ fs when the shock reflected ions leave the target. One can notice a reduction in the energy of the quasi-monoenergetic ions on accounting for RR force and the PP effects. One may also note that the number of particles of the accelerated bunch in the RR and RR+PP case is higher. The energy per carbon ion without radiation reaction is  $324$ MeV (27 MeV/u) with an energy spread ($\Delta E/E$) of about $4\%$ at FWHM (full-width at half-maximum), while accounting for the RR force and pair plasma formation, it drops to $192$ MeV ($\sim 16$ MeV$/$u) with $2.5\%$ FWHM , and $168$ MeV ($\sim 14$ MeV$/$u) with $2\%$ FWHM, respectively. The obtained energies are similar to ones obtained recently at lower laser intensity but with higher energy spread~\cite{Pak:2018aa}. The FWHM of the ion energy spectra become better on accounting for the RR force and PP production, which is extremely encouraging. This can be attributed to the smoother shock front formation as seen in Fig.\ref{fig1}. At later times (570 fs), as can be seen in the Fig.~\ref{fig_fe} (bottom panel) the ion energy gain increases while keeping the FWHM quite low, viz. 928 MeV (78 MeV/u) with $\sim 2.6\%$ spread, 495 MeV (41 MeV/u) with $\sim 2.8\%$ FWHM, 465 MeV (39 MeV/u) with $\sim 1.7\%$ FWHM, respectively.{At this instant, one can see a broader peak at lower energy in the case of no RR force.  Since the TNSA is present, the plasma target expands in vacuum at the rear-side. In this scenario, the ions have total velocity, $v_{ions} = 2 v_{sh} + v_{\textrm{exp}}$, where $v_{\textrm{exp}}$ is the plasma expansion velocity in vacuum. If $v_{\textrm{exp}} \ll v_{sh}$, then FWHM of the ion energy spectra is not severely affected. If both velocities are comparable \emph{i.e.} $v_{\textrm{exp}} \sim v_{sh}$, then the final ion energy also gets an additional boost due to the plasma expansion velocity in vacuum ($v_{ions} = 2 v_{sh} + v_{\textrm{exp}})$, resulting in higher ion energy gain albeit at the expense of degradation in the ion energy spectrum. In the case of $v_{\textrm{exp}} > v_{sh}$, the ion energy spectrum significantly broadens. This can be seen in the phase space of electrons and ions in Fig.\ref{fig_xpx}. TNSA ions at the back of the target have higher energies but a large energy spread. Though, the RR force and PP weaken TNSA (largely due to the reduction in the hot-electron energy), TNSA of ions is still dominant (see also insets in Fig.~\ref{fig_em}). In the no RR case, two ion energy peaks (solid black line in the lower panel of Fig.\ref{fig_fe}) corresponding to CSA (high-energy $\sim 1000$ MeV) and TNSA (low-energy $\sim 350$ MeV) of ions, can be clearly discerned. However, in the later two cases (RR and RR+PP), TNSA of ions causes merging with the shock accelerated ions and the ion energy peaks (blue and red lines in the lower panel of Fig.\ref{fig_fe}) start to become broader. Thus, TNSA of ions has an adverse impact on the ion energy spectrum. To minimize the impact of TNSA, one can resort to tailored target profile and we study it in Sec.\ref{tailored_target}.

\section{CSA of ions  for a tailored plasma target}\label{tailored_target}

As mentioned earlier, apart from employing the longer laser pulse durations, one can also further improve the energy and the spectrum of the beam by target engineering~\cite{Fiuza:2012ab,Fiuza:2013aa}. Tailored target with a slowly decreasing density profile at the back of the target reduces the sheath field to a small constant value, making it uniform. The late-time evolution of the ion energy spectrum is dominated by TNSA of ions which causes a broader Maxwellian spectra. The relative dominance of CSA and TNSA of ions is physically connected with the shock reflection time and the Debye sheath formation  at the back of the target. Consideration of these two aspects yields (without accounting for RR force and PP) the optimum target scale length, $l_{s} = (m_i/Z m_e)^{1/2} \lambda_0 /2$, where $m_i$ is the carbon ion mass and $\lambda_0=1\,\mu$m is the wavelength of the laser pulse~\cite{Fiuza:2013aa}. We carried out 2D PIC simulations with the same laser-plasma parameters, but now with a tailored plasma density profile. The laser pulse with normalised vector amplitude $a_0=300$ interacts with a pre-formed fully-ionised carbon plasma (C$^{6+}$) with a temperature $T_{e^-}=T_{\text{C}^{6+}}=700$ eV and a maximum electronic density, $n_{e}=300n_c$ with the following density profile
\noindent
\begin{equation}
n_e(x) =  300n_c \left\{
  \begin{tabular}{ll}
        $x/x_1$  & $x \leq x_1  \,$, \\
        $e^{-  (x-x_1) /d_{s}} $ & $x >x_1  \, $,
  \end{tabular}
\right.
\end{equation} 
where $x_1 = 5\,\mu$m, upto which the density increases linearly and then decays exponentially with a scale length of $\sim 10$ microns~\cite{Fiuza:2013aa}. We investigated few optimum scale lengths and show here the results for, $ d_s= l_s/3$, case. The simulation box has dimensions of $L_x \times L_y = (150\mu \textrm{m} \times 6\mu \textrm{m}$), with the cell size: $\Delta_x \times \Delta_y = (35 \textrm{nm} \times 35 \textrm{nm}$) and uses $50$ particles per cell. For this density profile, the laser pulse
 propagating in an underdense plasma region, $x < x_1$, strongly heats the plasma electrons which helps in launching an electrostatic shock at $x=x_1$. This shock then propagates in a plasma with the spatial decay of the density and accelerates the ions in the upstream region. However, due to decaying density profile 
 the plasma space-charge field becomes weaker (for $x > x_1$) and the laser radiation pressure marginally dominates over it, yielding higher piston and shock velocities, compared to the slab target case. This results in higher acceleration of ions. Consequently, the shock suffers strong dissipation as it further propagates in the plasma. Since the plasma density is also decaying, the number of the upstream ions reflected by the shock also becomes smaller, saturating CSA of ions. Moreover, at later times, the laser radiation pressure (if the laser pulse energy is not severely depleted) starts strongly dominating over the space charge field (due to lower plasma density), and accelerate the ions population closer to the shock front by the RPA mechanism. Thus, for the tailored targets, at later times, the CSA  of ions is further complimented by the RPA of ions resulting into two groups of high-energy ions. This significantly complicates the further target density optimisation for the stronger CSA of ions.

\subsection{Shock structure formation and electron-ion phase spaces}
 
\begin{figure}
\includegraphics[height=0.38\textheight,width=0.52\textwidth]{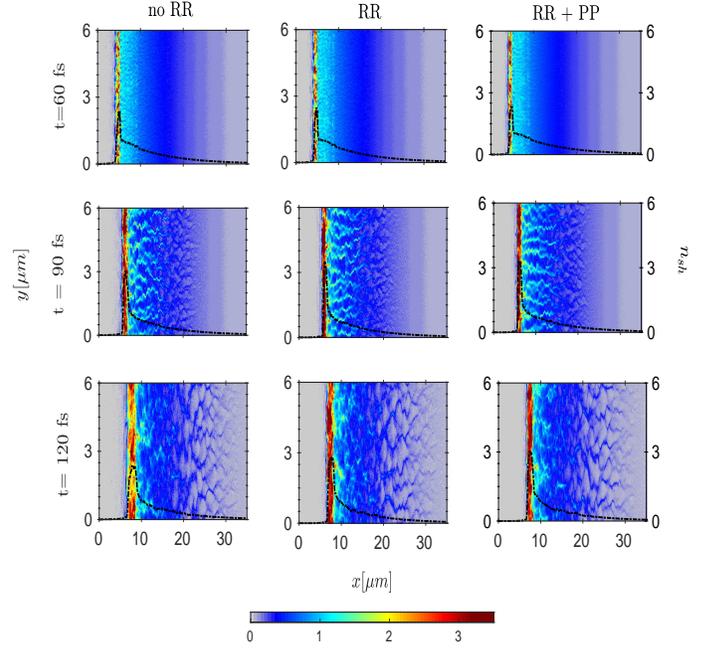}
\caption{Number density of plasma normalised by initial plasma density at different instants. First, second and third columns represent the cases of no RR, with RR and with RR + PP respectively. The shock density jump, averaged in the $y$-direction, is over-plotted (second $y$-axis) with a black dotted line in each case.}
\label{fig1t}
\end{figure}

\begin{figure}
\centering
\includegraphics[height=0.38\textheight,width=0.48\textwidth]{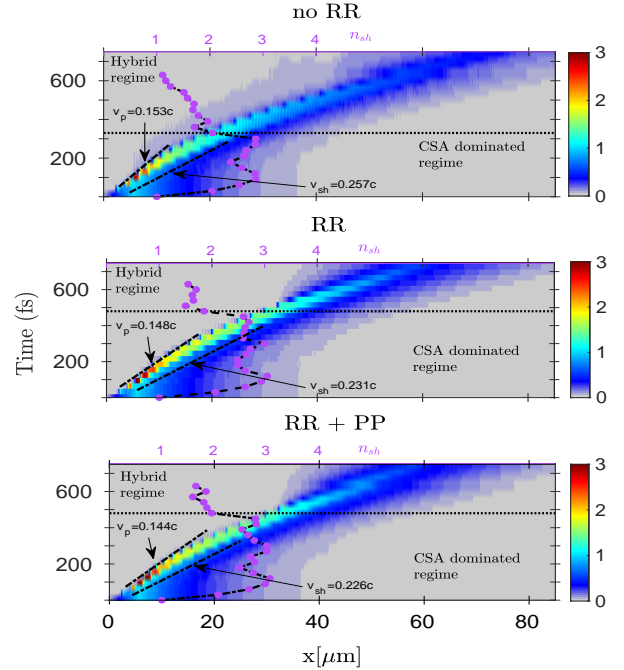}
\caption{$(x$-$t)$ plot of the ion density indicating shock and  piston velocities in the three cases considered. Also over-plotted is the shock density jump, $n_{sh}$ that clearly demarcates the CSA dominated regime from the RPA dominated. }
\label{fig_xt_tailored}
\end{figure} 

Fig.\ref{fig1t} shows the time evolution of tailored target density and one can see a stable shock structure formation in the case of RR (second column) and RR + PP (third column) as also in Fig.\ref{fig1}. However, the shock width is narrower on account of the peak target density being concentrated to a thinner region. The time evolution of the plasma density jump (normalized by the peak density $n_e=300 n_c$) is further depicted in Fig.\ref{fig_xt_tailored}. One can see that both the piston and shock velocities are higher compared to the slab target case in Sec.\ref{slab_target}. This is attributed to the marginal dominance of the laser radiation pressure over the space charge field as discussed before. We note here that the shock density jump is defined as $n_{sh}=n_d/n_u$, here $n_u$ is decaying with $x$ and hence can't taken as $n_u=300 n_c$ as in the slab target case. Thus, we overplot the density jump associated with the shock at each instant by using the local values of $n_d$ and $n_u$ in Fig.\ref{fig_xt_tailored}. Since the target density decays exponentially after $x >x_1$, the shock actually lasts longer than depicted via the contour plot in Fig.\ref{fig_xt_tailored}. Nevertheless, due to the higher shock velocity, it dissipates its energy to the ions faster and the density jump associated with it becomes smaller, $n_{sh} < 2$, at an instant marked by the horizontal dashed lines overplotted in each case in  Fig.\ref{fig_xt_tailored}. This horizontal line shows the time at which the ion acceleration enters the hybrid regime of CSA and RPA of ions.  At this instant the radiation pressure of the laser pulse dominates over the space-charge field and can accelerate the ions by the RPA mechanism. Thus, in the CSA regime, the shock density jump oscillates between $n_{sh}=(2-3)$. As the shock weakens, the density jump falls below $n_{sh} < 2$ and the ion acceleration enters the hybrid regime where the RPA mechanism begins to play an important role. The transition from CSA to RPA of ions is faster in the case of no RR force (first panel). 
\noindent 
\begin{figure}
\includegraphics[height=0.32\textheight,width=0.5\textwidth]{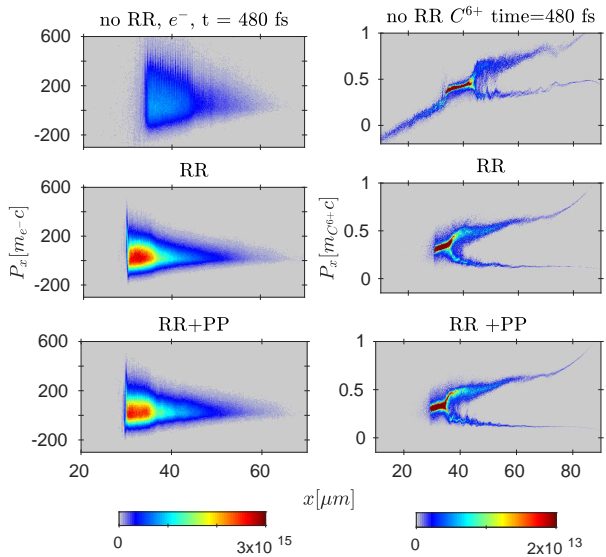}
\caption{Electron phase space (left column) and ion phase space (right column) at $t = 480 $ fs for all 3 cases.}
\label{fig_xpx_t}
\end{figure}
\noindent
Since in this case, the laser energy depletion into high-energy photons isn't accounted for, the laser radiation pressure is strong enough to start accelerating the ions earlier compared to the later two cases. In the last two cases (RR and RR+PP), the laser energy partitioning into high-energy photons and pairs depletes the laser energy considerably and the transition to RPA regime begins later compared to the first case (no RR). Though, the filamentary structures look qualitatively same as in Fig.\ref{fig1}, the hot electron transport, and consequently TNSA of ions, in this case differs from the tailored target case. This can be clearly seen by comparing the electron and ion phase spaces for the slab target case in Fig.\ref{fig_xpx} from Sec.\ref{slab_target} with that of tailored target case in Fig.\ref{fig_xpx_t}. One can immediately notice the suppressed TNSA of ions in Fig.\ref{fig_xpx_t} as expected~\cite{Fiuza:2012ab,Fiuza:2013aa}. This makes it possible to study the acceleration of ions at longer time scales as discussed in Sec.\ref{slab_spectrum}. One also notices that in the case of RR (also RR+PP), the relative suppression of TNSA of ions is stronger, as also seen in Fig.\ref{fig_xpx}. It is worthwhile to note here that in the case of no RR force, a strong heating of the hot-electrons occurs at early time and the shock is launched. After the shock is launched, the further generation of hot-electrons is not significant at later instants as seen in the upper panel of Fig.\ref{fig_xt_tailored}.

\subsection{Ion energy spectra}
\begin{figure}
\includegraphics[height=0.22\textheight,width=0.52\textwidth]{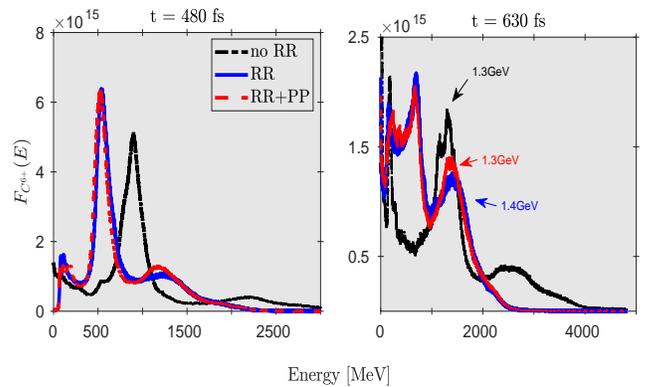}
\caption{Ion energy spectrum at $t=480$ fs. The case without the radiation reaction force (black line) has a peak energy of 900 MeV with an energy spread of $\sim 20\%$ while for case with RR and RR + PP it is 551 MeV with a spread of $\sim 25\%$ and 548 MeV with the same spread. At a later time, (630 fs) the energies gained are much higher for each case, but the energy spread broadens viz. 1.302 GeV ($\sim 28\%$), 1.432 GeV ($\sim 47\%$) and 1.382 GeV ($ 46\%$).}
\label{fig_fe_t}
\end{figure}

Fig.\ref{fig_fe_t} shows the evolution of ion energy spectrum for the three cases. Due to target tailoring, the TNSA of ions is not dominant, and is further suppressed in the case of the RR and RR+PP. At an earlier instant, (480 fs), the energy in the case of no radiation reaction force is 900 MeV ($\sim 75$ MeV$/$u) with an energy spread ($\Delta E/E$) of about $\sim 20\%$ FWHM, while with the RR force and pair plasma formation, it drops to 540 MeV ($\sim 45$ MeV$/$u) with $25\%$ FWHM for both the cases. It should be pointed out here that, the energy spread $\Delta E$ in the case without RR is larger ($\sim $190 MeV) than the other cases ($\sim $144 MeV each for RR and RR+PP), but since the peak energy ($\sim $ 900 MeV) is much larger than the other cases ($\sim $ 550  MeV), the percentage $(\Delta E/E) \%$ is relatively smaller. 
Though the ion energy spread of these ions is also larger than the previous case (Fig.\ref{fig_xpx}). The larger energy spread is attributed to the stronger shock dissipation that can cause non-uniform acceleration and hence larger energy-spread in the ion spectra.
Later on (at 630 fs), it can be seen that ions gain much larger energy $2.6$ GeV ($220$ MeV/u ) without the RR force, 1.4 GeV ($116$ MeV/u) with the RR force and 1.3 GeV ($108$ MeV/u ) in the case of (RR + PP). An interesting consequence of the interplay between the RPA and CSA of ions can be seen at $t=630$ fs, where one notices two peaks formation in the ion energy spectra. The RPA of ions, in the later phase, can accelerate a large number of ions but due to laser energy depletion and the decaying plasma density is not very effective in accelerating the ions to higher energies. The second group of ions accelerated by the collisionless shock has higher energies since the CSA of ions has occurred on a longer time scale. Due to higher shock velocity in the no RR case, the shock accelerated peak with energy 2.6 GeV disappears on longer time scales. But a second peak (presumably due to the RPA mechanism) that is around $1.3$ GeV survives. Hence, on longer time scales, accounting for the radiation losses and plasma pair formation lead to a higher energy gain, which is indeed  encouraging. The ion energy peaks at longer times have the energies, 1.302 GeV (108 MeV/u) with $28\%$ FWHM without RR, 1.432 GeV (119 MeV/u) with $47\%$ FWHM with RR, and 1.38 GeV (115 MeV/u) with $46\%$ FWHM with (RR + PP) case.  
\section{Conclusion}

To conclude, for the first time, the CSA of the ions in ultra-relativistic regime of the laser-plasma interaction has been studied by allowing, ab-initio, accounting of the RR force and pair-plasma formation in a carbon-electron plasma. Accounting for the radiation reaction force and pair production result in lower shock velocities and consequently lower ion energy $(\sim 50)$ MeV/u for a slab target case. Nevertheless, the energy spread is rather small $\sim 2$-$3\%$ in the ultra-relativistic regime where the effect of the RR force and PP formation is important. The narrow energy spread is attributed to the smoother shock front formation in this case. By employing the tailored target, one can achieve higher ion energy gain ($\sim 120$ MeV/u) albeit the energy spread becomes higher \emph{e.g.} $\sim 30\%$, presumably due to the strong shock dissipation.  Both the longer laser pulse duration and the optimised target density profile yield higher energy ion energy.  The formation of the ion energy peak at longer time duration is favourable for experimental realization of the scheme. For the intensities considered in this paper, one can get carbon ions with $\sim 120$ MeV/u, which matches the minimum ion energy required for the use of carbon ions in tumour therapy.  These results can be further improved by employing the longer laser pulses and optimizing the density scale length at the back of the target. Thus, the higher ion energies obtained in a tailored target case and the lower energy spread in a slab target case represent a significant step forward in this direction.

\acknowledgments{We thank Prof. Christoph Keitel for reading the manuscript and making a number of useful comments.}

%

\end{document}